\documentclass{article}
\usepackage{spconf,amsmath,amssymb,graphicx,hyperref}
\hypersetup{hypertexnames=false}  
\usepackage{float,multirow}


\title{Taming Long-form Text-to-Speech}

\name{Rongxiang Wang$^{1,2}$\sthanks{Work done during an internship at Argmax, Inc. Corresponding author: waq9hw@virginia.edu}, Berkin Durmu\c{s}$^{1}$, Ay\c{s}eg\"ul Orhon$^{1,3}$, Eduardo Pacheco$^{1}$, Atila Orhon$^{1}$}
\address{$^{1}$Argmax, Inc. \quad $^{2}$University of Virginia \quad $^{3}$Bilkent University}

\makeatletter
\g@addto@macro\@maketitle{%
  \vspace{1.0em}%
  \centerline{\includegraphics[width=0.92\textwidth]{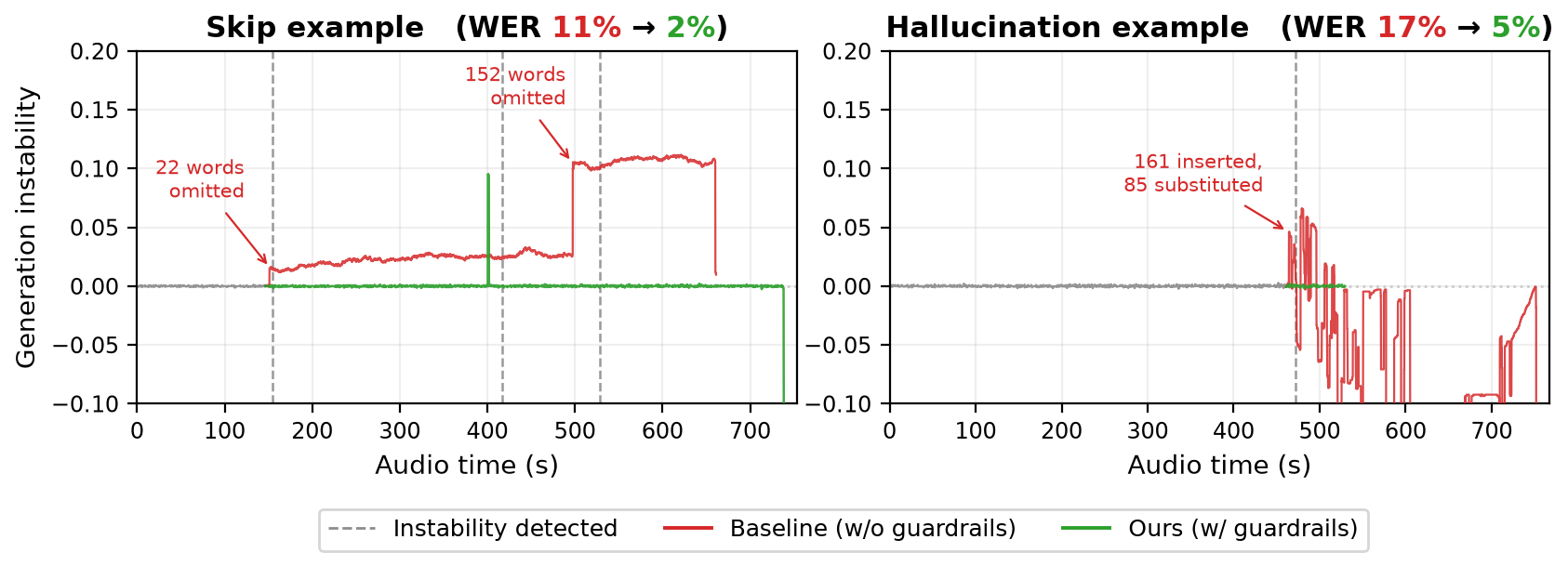}}%
  \refstepcounter{figure}\label{fig:traj}%
  \vspace{0.3em}\par\noindent
  \begin{minipage}{\textwidth}\footnotesize\textbf{Fig.~\thefigure}.\ \textbf{
   LACI detects and repairs long-form generation failures.}
   The plots show LACI's instability measure during long-form generation with Qwen3-TTS-0.6B-CustomVoice on a multi-minute customer support call session from~\cite{apptek_callcenter}.
   Skip and hallucination, the two failure types we define, appear in the red trajectories. LACI recovers from both, as shown by the green trajectories.\end{minipage}%
  \vspace{0.8em}%
}
\makeatother

\begin{document}
\ninept
\maketitle

\begin{abstract}
Long-form text-to-speech (TTS) enables multi-turn conversations with consistent prosody and higher
quality voice cloning from longer reference audio. Recent open-weights autoregressive TTS models
such as Qwen3-TTS and VoxCPM2 attain state-of-the-art word error rate (WER) and speaker similarity
(SIM) on short-form prompts but significantly deteriorate when used with long-form prompts. We
propose Localized Attention-Constrained Inference (LACI), an inference-only method to detect TTS errors
in near real-time, roll back to the error onset and regenerate with
temporary guardrails, adding negligible computational overhead. Using LACI, we improve worst-of-$N$
WER across 10 RNG seeds for Qwen3-TTS-0.6B from 35.2\% to 3.4\% on prompts longer than 1500 words, even
surpassing its short-form reliability of 5.4\% on prompts with fewer than 500 words. To demonstrate the
efficacy of LACI on voice cloning reliability, we propose a sliding-window version of the SIM metric
that we call wSIM. wSIM exposes several novel failure patterns that are not captured by SIM. LACI
improves worst-of-$N$ wSIM from 0.01 to 0.47 on 120 seconds of reference audio while reducing the
rate of catastrophic generations with WER above 30\% from 26\% to below 1\%.
\end{abstract}


\begin{figure*}[t!]
   \centering
   \includegraphics[width=0.9\textwidth]{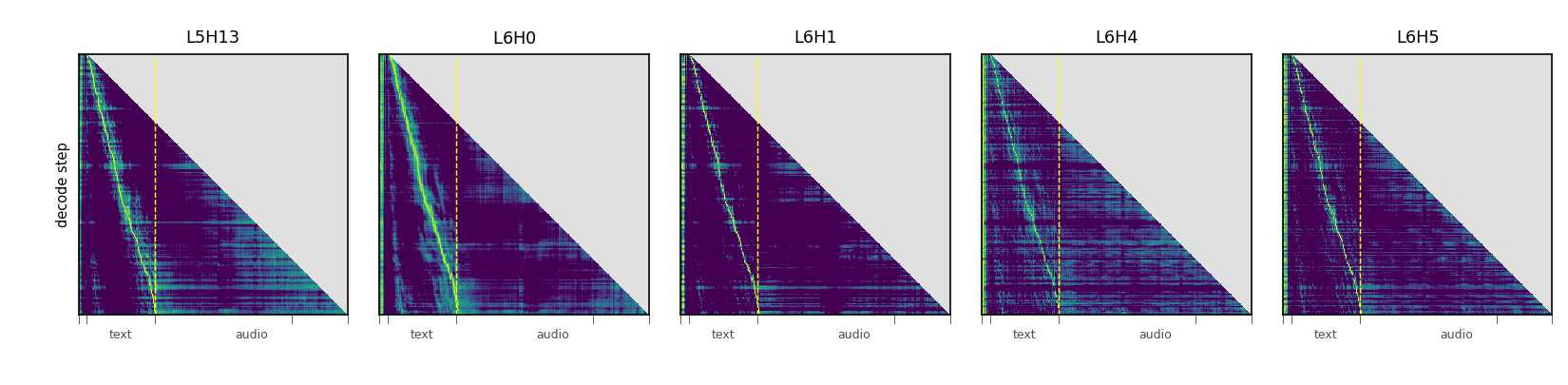}
   \caption{\textbf{Audio-text alignment heads in autoregressive neural codec language models.}
   All variants of Qwen3-TTS and VoxCPM2 have emergent \textit{alignment heads} that map corresponding audio and text tokens.
   Shown are the attention heatmaps of Qwen3-TTS-0.6B-CustomVoice's alignment heads, which are consistent across text and speaker prompts.}
   \label{fig:headtax}
\end{figure*}

\section{Introduction}
\label{sec:intro}

Autoregressive neural codec language models power many recent state-of-the-art
text-to-speech (TTS) systems~\cite{valle,valle2,qwen3tts}. As these systems graduate from
academic benchmarks to audiobook narration, voice agents for customer support and voice cloning from
multi-minute reference audio, reliably generating speech from long-form text prompts becomes essential
for consistent voice timbre and prosody.

While Qwen3-TTS and VoxCPM2 excel at short-form generation, their reliability
deteriorates as the input text grows, leading to high mean and variance in content accuracy as measured
by word error rate (WER). For voice cloning, this means that they cannot benefit from long reference audio to improve speaker similarity. In this paper:

\begin{itemize}
   \itemsep0.1em
      \item We show that the reliability gap between short- and long-form generation is attributable to attention misalignment between generated
 audio and input text tokens for Qwen3-TTS-0.6B, Qwen3-TTS-1.7B and VoxCPM2.
      \item We define two failure types that explain this gap and propose an efficient failure detector based solely on the system's attention weights during streaming generation.
      \item We propose \textbf{Localized Attention-Constrained Inference (LACI)}, which rolls back to the detected failure onset and imposes temporary guardrails on attention weights to recover.
      LACI removes the gap between short- and long-form generation and generalizes across model sizes (Qwen3-TTS-0.6B and 1.7B) and model families (VoxCPM2).
\end{itemize}

\section{Related Work}
\label{sec:related work}

\noindent \textbf{Training-time methods}. Traditional TTS architectures control attention alignment during \emph{training}.
Tacotron 2~\cite{tacotron2} uses location-sensitive attention to bias alignment toward monotonic progress.
Duration models and forced-alignment objectives supply an explicit text-audio map~\cite{onealignment}, and remain standard in compact non-autoregressive systems such as Kokoro~\cite{kokoro}.
Modern neural codec language models adopted similar objectives~\cite{ralle,valler,oas}. All of these require training-time intervention and may impose challenges to training stability.

\noindent \textbf{Inference-time methods}. One class of inference-only methods generates multiple candidates from different RNG seeds and picks the
one with the lowest WER according to an external speech recognition system~\cite{seedtts,llasa}. The limitation of this approach is that the inference cost scales linearly with
the number of candidates, and the method does not attempt to fix a failure which might persist across different RNG seeds.
An ideal solution should instead detect and recover from the failure in near real-time at negligible computational overhead.
The closest attempt is Attention-Constrained Inference (ACI)~\cite{aci}, developed on earlier decoder-only TTS systems, which
enforces an \emph{always-on} hard attention mask on selected attention heads at every generation step.
We adapt ACI to Qwen3-TTS and VoxCPM2 and show that it narrows the reliability gap between short- and long-form generation but does not close thegap completely.
Furthermore, because it is always on, it introduces new errors where the baseline was correct, and because it lacks failure detection, it
is unable to address persistent failures that require multiple attempts to recover from.
Our method, \textbf{LACI}, improves ACI from an always-on guardrail into one that detects failures, rolls back and guards. LACI addresses all three of the aforementioned limitations.

\begin{figure*}[t!]
   \centering
   \includegraphics[width=0.9\textwidth]{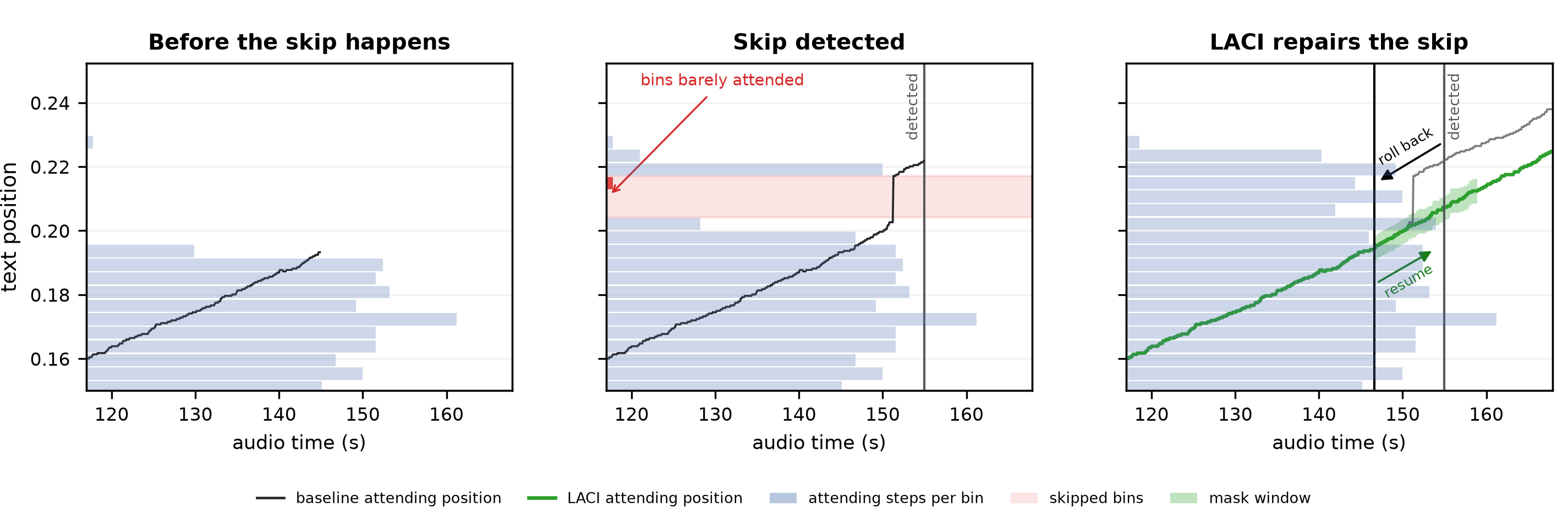}
   \caption{
      \textbf{Detecting and recovering from a \textit{skip} failure.}
      At each generation step the system attends to one text position, the argmax over attention weights.
      LACI monitors cumulative attention coverage across bins of input text tokens.
      When a bin crosses the coverage threshold before the bins before it do, a skip is detected.
      LACI rolls back and temporarily applies a hard attention mask to regularize attention coverage and recover.
   }
   \label{fig:mechanism}
\end{figure*}

\section{Method}
\label{sec:method}

\subsection{Failure Definition}

Content accuracy degrades when a TTS system substitutes, inserts or deletes words while mapping input text to speech. Qwen3-TTS-0.6B achieves 5.4\% worst-of-$N$ WER across 10 RNG seeds on prompts with fewer than 500 words.
This number increases to 35.2\% when running the same system on the same dataset with concatenated prompts that are longer than 1500 words.
Analyzing this gap, we identify two dominant failure modes: \textit{skip} and \textit{hallucination}. A skip is a contiguous deletion of 10 or more words, and a hallucination is a contiguous mix of substitutions and insertions of 20 or more words. Figure~\ref{fig:traj} shows an example of each.
Other errors occur at the same rate in short- and long-form generation and set the content accuracy upper bound of a model without further training or inference harness improvements.

\subsection{Failure Detection}
Analyzing the attention weights of Qwen3-TTS-0.6B, we identified several heads with emergent text-audio alignment during generation sessions with high content accuracy (Figure~\ref{fig:headtax}).
Their alignment pattern is consistent across text prompts and speakers during high content accuracy sessions, and the alignment breaks during sessions with low content accuracy.
We use the same analysis to identify alignment heads in Qwen3-TTS-1.7B and VoxCPM2 as well.
Figure~\ref{fig:traj} summarizes this signal as a single instability measure, the deviation of the alignment head's reading position from its expected pace. It stays at zero while the model reads on pace and departs from zero at the onset of a skip or hallucination.

Equipped with these observations, we build a detector that flags a skip or hallucination failure within just a few seconds of output audio.
The detector splits the input text prompt into equal-sized token bins and counts how many generation steps the alignment head has spent attending to each bin.
When the reading position moves past a token bin that received far fewer steps than the bins before it, the detector flags a skip failure.
When it stops advancing for much longer than a bin normally takes, the detector flags a hallucination failure.
Once the last bin has received its share of attention, the text has been fully read and the detector ends the generation.
Figure~\ref{fig:mechanism} shows a skip detected within a few seconds of output audio after its onset.
The instability measure, the detector settings and the recovery procedure are documented in our code, which will be open-sourced upon publication.

We report detection latency in seconds of generated audio because it is independent of hardware.
The wall-clock cost is this figure divided by the real-time factor of the inference system.
Leading implementations of Qwen3-TTS generate several times faster than real time and stream into a playback buffer, so the buffered audio covers detection, rollback and regeneration without interrupting playback.

\subsection{Failure Recovery}

Near-real-time detection makes it possible to recover without regenerating from scratch.
LACI rolls back to the onset of the detected failure, reseeds the RNG and applies a hard attention mask that enforces text-audio alignment until generation moves past the point of detection.
The mask is the same as in ACI~\cite{aci}, but it is applied only during this probationary period.
Persistent failures may trigger LACI several times so we apply a cap on maximum retries to bound the computational overhead.

\section{Experiments}
\label{sec:experiments}

We evaluate on the AppTek Call Center dataset~\cite{apptek_callcenter}, a real-world long-form TTS use case.
Starting from full customer support call sessions with agent-side-only audio and transcripts, we extend coverage to shorter and longer prompts
while keeping the domain and prompt difficulty constant (Table~\ref{tab:corpus}).
Speech turns represent a voice agent whose TTS component only sees a single response, which establishes short-form reliability but may sound inconsistent across turns.
Full sessions represent a TTS component with the context of an entire session, and multi-session prompts one that also carries past sessions for maximum speaker consistency.

\begin{table}[t]
   \centering
   \small
   \setlength{\tabcolsep}{4pt}
   \begin{tabular}{lrrl}
   \hline
   Prompt & Words & Reference audio  \\
   \hline
   Speech turn & 15--316 & 0.1--2 min \\
   Full session & 331--1579 & 2--9 min \\
   Multi-session & 1349--2449 & 9--14 min \\
   \hline
   \end{tabular}
   \caption{\textbf{Prompt length buckets from the AppTek Call Center dataset.}
   The 3 splits keep the domain and prompt difficulty fixed while covering short-form and long-form generation scenarios.}
   \label{tab:corpus}
\end{table}

We curate 144 prompts across these buckets and evaluate each across 10 RNG seeds per TTS system. Evaluation across many seeds measures the reliability of these sampling-based systems. Specifically, we report worst-of-$N$ WER in addition to mean WER to quantify the latent reliability of a system when prompt difficulty is controlled exactly.
We evaluate Qwen3-TTS-0.6B-CustomVoice, Qwen3-TTS-1.7B-CustomVoice and VoxCPM2.

For voice cloning, we start from a single speaker turn of a session as the shortest reference and build references up to 120 seconds, while fixing the text to generate as the last minute of each session, across 10 RNG seeds. We use Qwen3-TTS-0.6B-Base for these experiments.

Content accuracy is the WER between the NVIDIA Parakeet-TDT-0.6B-v2~\cite{parakeet} transcription of the generated audio and the ground truth prompt, after text normalization following~\cite{whisper}. Speaker similarity (SIM) is the cosine similarity between speaker embeddings from WavLM-Large fine-tuned for speaker verification, as in Seed-TTS~\cite{seedtts}.
We also introduce wSIM, a sliding-window version of SIM (8-second windows, 4-second stride) that exposes failure patterns SIM misses, such as short spans of speaker identity instability within a long generation. We report worst-of-$N$ wSIM, the lowest wSIM across all windows and RNG seeds for a given reference and prompt. We manually verified that the lowest wSIM scores correspond to actual speaker identity switches or severe distortions.


\begin{table}[t]
\centering
\small
\setlength{\tabcolsep}{4pt}
\begin{tabular}{lccc}
\hline
Prompt length (words) & Baseline & ACI~\cite{aci} & LACI (Ours) \\
\hline
$<$500 &  2.7/5.4  & 2.7/6.0 & \textbf{2.7/5.4} \\
500--1000 &  3.5/6.5  & 3.4/9.3  & \textbf{2.9/3.9} \\
1000--1500 &  4.2/11.2 & 2.8/6.0 & \textbf{2.5/3.3} \\
$\geq$1500 &  16.6/35.2  & 3.7/11.5 & \textbf{2.9/3.4} \\
\hline
\end{tabular}
\caption{\textbf{Qwen3-TTS-0.6B mean/worst-of-$N$ WER (\%) on the AppTek Call Center dataset ($N{=}10$ RNG seeds).}
Baseline is the reference Qwen3-TTS implementation~\cite{qwen3tts_code} without guardrails.
ACI narrows the long-form gap but degrades short-form reliability.
LACI improves mean and worst-case reliability at every length, even beyond the baseline's short-form results. Bold marks the lowest worst-of-$N$ WER per row.
}
\label{tab:main06b}
\end{table}

\begin{figure}[t]
  \centering
  \includegraphics[width=0.90\columnwidth]{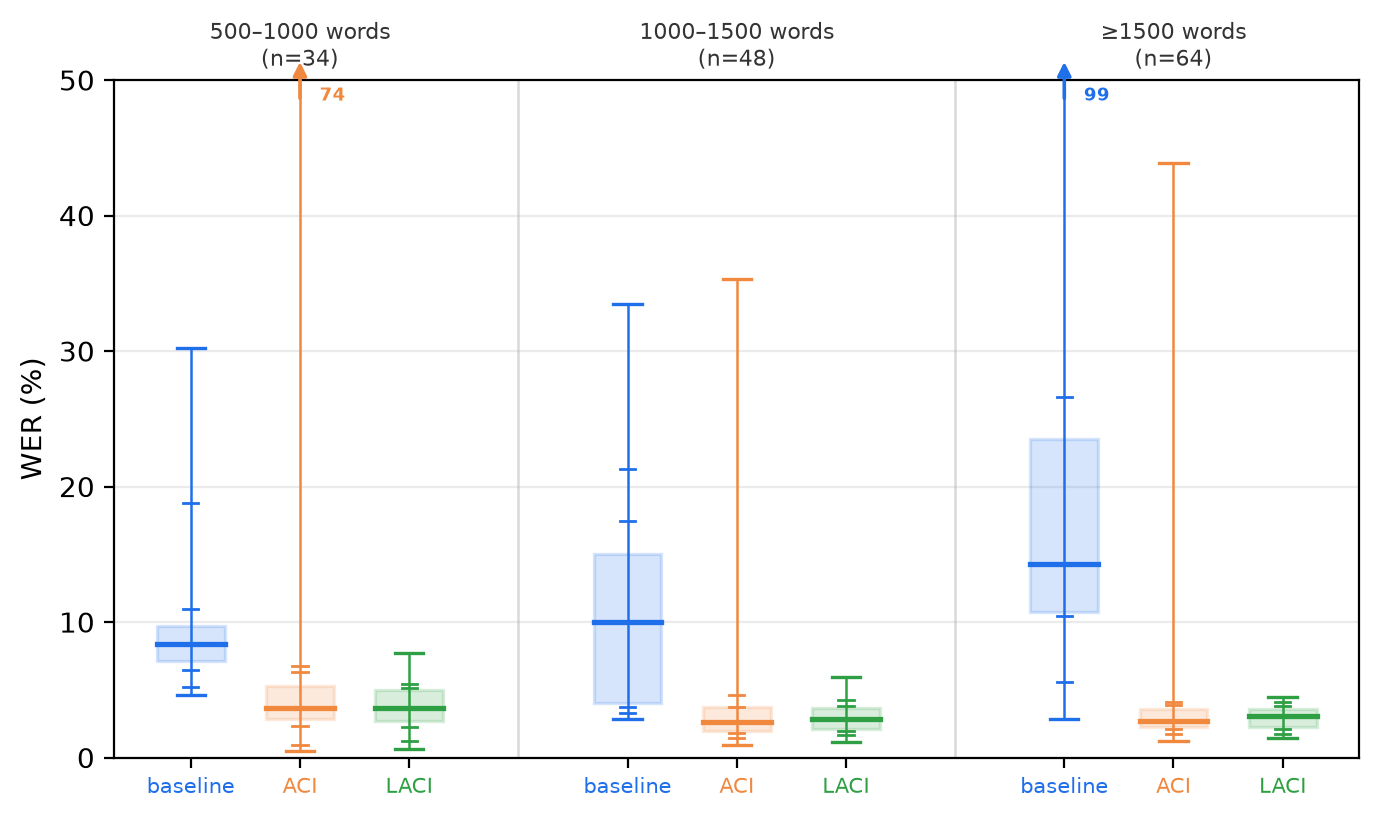}
  \caption{\textbf{LACI repairs the baseline's failures without leaving a tail.}
  WER of the three systems on the 146 (prompt, RNG seed) pairs where the baseline has at least 1 skip or hallucination failure, grouped by prompt length. Boxes show the median and interquartile range, whiskers the minimum to maximum (clipped at 50\%, true maximum labeled), ticks the 10th, 20th, 80th and 90th percentiles.
  ACI lowers the median but keeps a high upper whisker, while LACI brings every failure into a low, tight band.}
  \label{fig:boxlong}
\end{figure}

\section{Results}
\label{sec:results}

\noindent \textbf{Long-form reliability}. Table~\ref{tab:main06b} reports mean and worst-of-$N$ WER for Qwen3-TTS-0.6B. The baseline is reliable on prompts with fewer than 500 words but degrades steadily with prompt length, and its worst-of-$N$ WER reaches 35.2\% beyond 1500 words, more than six times its short-form value.
LACI holds the mean between 2.5\% and 2.9\% and the worst-of-$N$ WER between 3.3\% and 5.4\% at every length. The worst seed of a prompt longer than 1500 words is now more reliable than the worst seed of a short prompt without guardrails.
Figure~\ref{fig:boxlong} isolates the 146 (prompt, RNG seed) pairs where the baseline has at least 1 skip or hallucination failure. LACI brings every one of them into the same low band as short prompts, whereas ACI lowers the median but leaves a long upper tail.

\noindent \textbf{Failure detection versus always-on guardrails}. ACI narrows the long-form gap and cuts worst-of-$N$ WER beyond 1500 words from 35.2\% to 11.5\%. However, its always-on mask disrupts generations that the baseline gets right, so worst-of-$N$ WER rises from 5.4\% to 6.0\% on short prompts and from 6.5\% to 9.3\% on prompts of 500 to 1000 words.
LACI intervenes only after a failure is detected. It matches the baseline exactly on short prompts and matches or improves on ACI at every prompt length.

\noindent \textbf{Generalization}. Table~\ref{tab:main17b} shows the same picture at a larger model scale and for another model family.
Qwen3-TTS-1.7B is less reliable than the 0.6B model on long prompts, with baseline worst-of-$N$ WER of 45.7\% and 47.8\% in the two longest buckets. LACI brings both below 4.5\%, while ACI still leaves 22.7\% on the longest prompts.
On VoxCPM2, LACI reduces the fraction of (prompt, RNG seed) pairs with a skip or hallucination in the 1000 to 1300 word range from 15.3\% to 3.3\%, and ACI reduces it to 2.0\%. Beyond 1300 words the failures resist regeneration. LACI lowers this fraction from 70\% to 57\% and ACI to 34\%, but neither repairs the tail, which points to a coherence limit of the model itself rather than a detection gap.

\begin{figure}[t!]
\centering
\includegraphics[width=0.77\columnwidth]{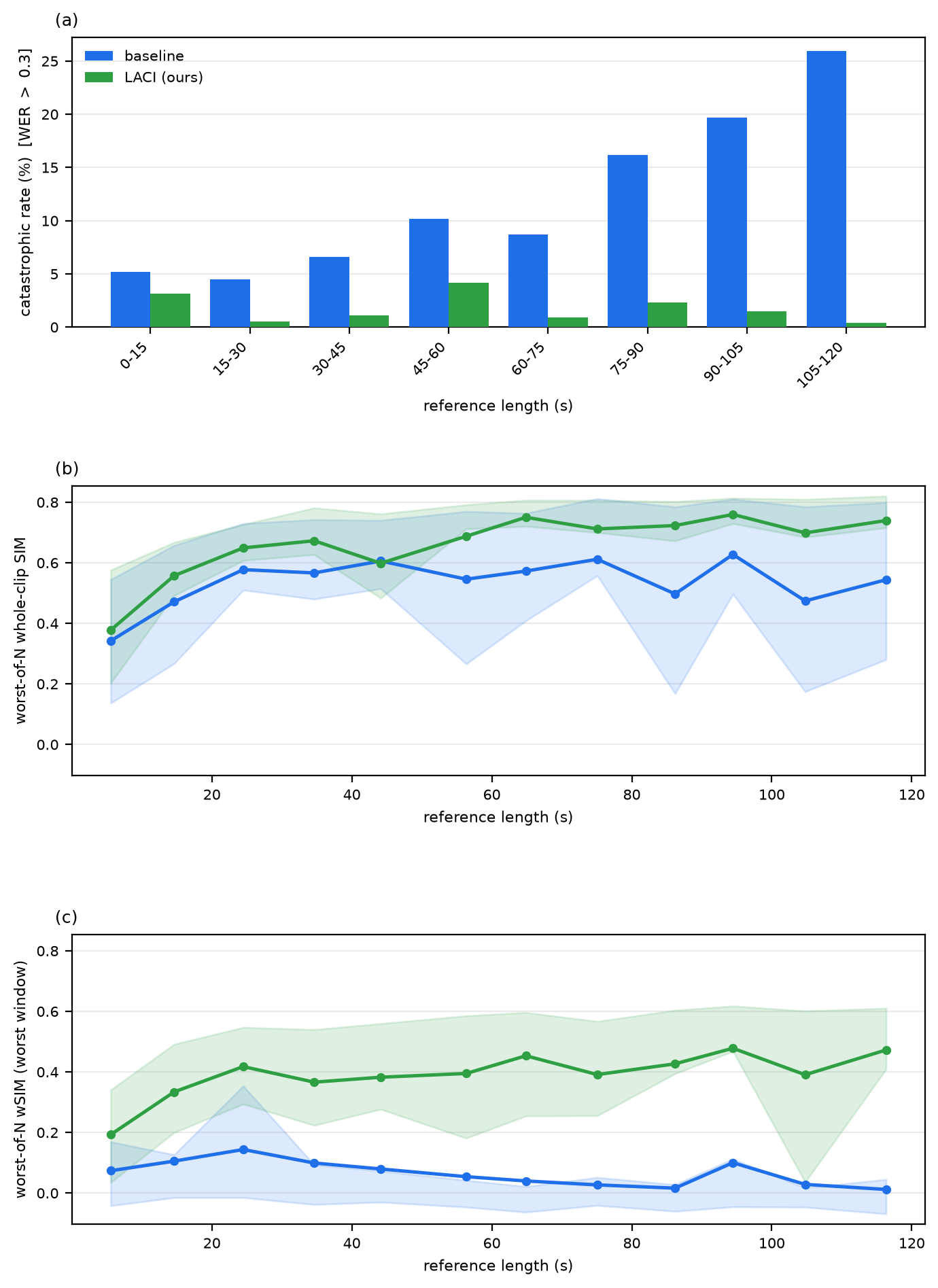}
\caption{
   \textbf{Qwen3-TTS-0.6B voice cloning reliability vs. reference audio length.}
   \textbf{(a)} Catastrophic failure rate (WER above 30\%) grows from 5\% to 26\% as reference audio grows to 120 seconds, while LACI keeps it below 5\%.
   \textbf{(b)} Worst-of-$N$ SIM across 10 RNG seeds improves with LACI.
   \textbf{(c)} Worst-of-$N$ wSIM collapses for the baseline while LACI holds it high.
}
\label{fig:swift}
\end{figure}

\noindent \textbf{Voice cloning}. Longer reference audio should give a voice cloning system more evidence about the target speaker, yet the baseline cannot benefit from it (Figure~\ref{fig:swift}).
Its catastrophic failure rate grows from 5\% with references shorter than 15 seconds to 26\% with references near 120 seconds. LACI keeps it below 5\% at every reference length and below 1\% at the longest.
While LACI improves on the traditional whole-clip SIM metric, this improvement is an understatement because a short span in which the voice switches identity or distorts is averaged out and hidden over a long clip. wSIM exposes these spans.
The baseline's worst-of-$N$ wSIM collapses to 0.01 at 120 seconds, while LACI reaches 0.47 and also improves whole-clip SIM.
With LACI, longer reference audio translates into higher speaker similarity instead of higher failure rates, and unlocks reliable and higher quality long-form voice cloning.

\begin{table}[t]
\centering
\small
\setlength{\tabcolsep}{2.3pt}
\begin{tabular}{llccc}
\hline
Model & Words & Baseline & ACI~\cite{aci} & LACI (Ours) \\
\hline
\multirow{4}{*}{Qwen3-TTS-1.7B}
 & $<$500  & 2.6/5.0 & 2.7/5.3 & \textbf{2.6/4.7} \\
 & 500--1000  & 4.5/11.9 & 3.5/10.2 & \textbf{2.8/3.8} \\
 & 1000--1500  & 15.6/45.7 & 2.8/8.5 & \textbf{2.3/3.0} \\
 & $\geq$1500 & 29.9/47.8 & 9.4/22.7 & \textbf{2.8/4.2} \\
\hline
\multirow{4}{*}{VoxCPM2}
 & $<$500 & 2.8/5.4 & 3.0/6.3 & \textbf{2.8/5.2} \\
 & 500--1000 & 3.4/5.2 & 3.5/5.2 & \textbf{3.4/4.4} \\
 & 1000--1300 & 5.2/16.3 & \textbf{3.0/4.3} & 3.5/8.1 \\
 & $\geq$1300 & 18.1/35.0 & 10.0/36.6 & \textbf{15.9/32.3} \\
\hline
\end{tabular}
\caption{\textbf{Qwen3-TTS-1.7B and VoxCPM2 mean/worst-of-$N$ WER (\%).} Same setup and bold rule as Table~\ref{tab:main06b}.
LACI generalizes to the larger Qwen3-TTS-1.7B with similar improvements and to VoxCPM2, where both guardrails repair the 1000 to 1300 word range.
Beyond 1300 words neither guardrail fully repairs VoxCPM2, indicating a model-coherence limit rather than a detection gap.
}
\label{tab:main17b}
\end{table}

\section{Future Work}
\label{sec:future}
LACI can be combined with a reinforcement learning environment where a TTS model can be used to generate many rollouts some with LACI and some without it.
This can be leveraged to efficiently generate many high and low quality text-to-speech generation.
TTS models trained on this data may exhibit higher reliability.
We invite other researchers to explore this direction.

\section{Conclusion}
\label{sec:conclusion}
We showed that the long-form reliability gap of state-of-the-art autoregressive TTS models is explained by audio-text misalignment on emergent alignment heads.
LACI monitors these heads to detect skips and hallucinations within seconds of output audio, and repairs them by rolling back and regenerating under temporary guardrails at negligible overhead.
This closes the long-form reliability gap for Qwen3-TTS-0.6B, carries over to Qwen3-TTS-1.7B and VoxCPM2, and lets voice cloning benefit from long reference audio instead of failing on it.

\vfill\pagebreak
\section{Acknowledgments}
\label{sec:ack}
This work was funded by Argmax, Inc. All authors are employees or interns of Argmax, Inc. The authors have no other relevant financial or non-financial interests to disclose.

\section{Compliance with Ethical Standards}
\label{sec:ethics}
This study uses only the publicly released AppTek Call Center dataset~\cite{apptek_callcenter} under its terms and collects no new human-subjects data, so no ethical approval was required.

\bibliographystyle{IEEEbib}
\bibliography{main}

\begin{thebibliography}{10}

\bibitem{apptek_callcenter}
Eugen Beck, Sarah Beranek, Uma Moothiringote, Daniel Mann, Wilfried Michel,
  Katie Nguyen, and Taylor Tragemann,
\newblock ``{AppTek Call-Center Dialogues: A Multi-Accent Long-Form Benchmark
  for English ASR},'' 2026,
\newblock arXiv:2604.27543.

\bibitem{valle}
Chengyi Wang, Sanyuan Chen, Yu~Wu, Ziqiang Zhang, Long Zhou, Shujie Liu, Zhuo
  Chen, Yanqing Liu, Huaming Wang, Jinyu Li, Lei He, Sheng Zhao, and Furu Wei,
\newblock ``Neural codec language models are zero-shot text to speech
  synthesizers,'' 2023,
\newblock arXiv:2301.02111 (VALL-E).

\bibitem{valle2}
Sanyuan Chen, Shujie Liu, Long Zhou, Yanqing Liu, Xu~Tan, Jinyu Li, Sheng Zhao,
  Yao Qian, and Furu Wei,
\newblock ``{VALL-E 2}: Neural codec language models are human parity zero-shot
  text to speech synthesizers,'' 2024,
\newblock arXiv:2406.05370.

\bibitem{qwen3tts}
Hangrui Hu, Xinfa Zhu, Ting He, Dake Guo, Bin Zhang, Xiong Wang, Zhifang Guo,
  Ziyue Jiang, Hongkun Hao, Zishan Guo, Xinyu Zhang, Pei Zhang, Baosong Yang,
  Jin Xu, Jingren Zhou, and Junyang Lin,
\newblock ``{Qwen3-TTS} technical report,'' 2026,
\newblock arXiv:2601.15621.

\bibitem{tacotron2}
Jonathan Shen, Ruoming Pang, Ron~J. Weiss, Mike Schuster, Navdeep Jaitly,
  Zongheng Yang, Zhifeng Chen, Yu~Zhang, Yuxuan Wang, R.~J. Skerry-Ryan, Rif~A.
  Saurous, Yannis Agiomyrgiannakis, and Yonghui Wu,
\newblock ``Natural {TTS} synthesis by conditioning {WaveNet} on mel
  spectrogram predictions,''
\newblock in {\em Proc. IEEE Int. Conf. Acoust., Speech, Signal Process.
  (ICASSP)}, 2018,
\newblock arXiv:1712.05884.

\bibitem{onealignment}
Rohan Badlani, Adrian {\L}a{\'n}cucki, Kevin~J. Shih, Rafael Valle, Wei Ping,
  and Bryan Catanzaro,
\newblock ``One {TTS} alignment to rule them all,''
\newblock in {\em Proc. IEEE Int. Conf. Acoust., Speech, Signal Process.
  (ICASSP)}, 2022,
\newblock arXiv:2108.10447.

\bibitem{kokoro}
{hexgrad},
\newblock ``{Kokoro-82M},'' \url{https://huggingface.co/hexgrad/Kokoro-82M},
  2025,
\newblock Hugging Face model card.

\bibitem{ralle}
Detai Xin, Xu~Tan, Kai Shen, Zeqian Ju, Dongchao Yang, Yuancheng Wang,
  Shinnosuke Takamichi, Hiroshi Saruwatari, Shujie Liu, Jinyu Li, and Sheng
  Zhao,
\newblock ``{RALL-E}: Robust codec language modeling with chain-of-thought
  prompting for text-to-speech synthesis,'' 2024,
\newblock arXiv:2404.03204.

\bibitem{valler}
Bing Han, Long Zhou, Shujie Liu, and Sanyuan Chen,
\newblock ``{VALL-E R}: Robust and efficient zero-shot text-to-speech synthesis
  via monotonic alignment,'' 2024,
\newblock arXiv:2406.07855.

\bibitem{oas}
Shiming Wang, Zhihao Du, Yang Xiang, Tianyu Zhao, Han Zhao, Qian Chen, Xiangang
  Li, Hanjie Guo, and Zhenhua Ling,
\newblock ``Eliminating stability hallucinations in {LLM}-based {TTS} models
  via attention guidance,'' 2025,
\newblock arXiv:2509.19852v1.

\bibitem{seedtts}
Philip Anastassiou, Jiawei Chen, Jitong Chen, Yuanzhe Chen, Zhuo Chen, Ziyi
  Chen, et~al.,
\newblock ``{Seed-TTS}: A family of high-quality versatile speech generation
  models,'' 2024,
\newblock arXiv:2406.02430.

\bibitem{llasa}
Zhen Ye, Xinfa Zhu, Chi-Min Chan, Xinsheng Wang, Xu~Tan, Jiahe Lei, et~al.,
\newblock ``Llasa: Scaling train-time and inference-time compute for
  llama-based speech synthesis,'' 2025,
\newblock arXiv:2502.04128.

\bibitem{aci}
Hankun Wang, Chenpeng Du, Yiwei Guo, Shuai Wang, Xie Chen, and Kai Yu,
\newblock ``Attention-constrained inference for robust decoder-only
  text-to-speech,''
\newblock in {\em Proc. IEEE Spoken Language Technology Workshop (SLT)}, 2024,
\newblock arXiv:2404.19723.

\bibitem{parakeet}
{NVIDIA},
\newblock ``{Parakeet-TDT-0.6B-v2},''
  \url{https://huggingface.co/nvidia/parakeet-tdt-0.6b-v2}, 2025,
\newblock Hugging Face model card.

\bibitem{whisper}
Alec Radford, Jong~Wook Kim, Tao Xu, Greg Brockman, Christine McLeavey, and
  Ilya Sutskever,
\newblock ``Robust speech recognition via large-scale weak supervision,'' 2022,
\newblock arXiv:2212.04356.

\bibitem{qwen3tts_code}
{Qwen Team},
\newblock ``{Qwen3-TTS}: Official repository,''
  \url{https://github.com/QwenLM/Qwen3-TTS}, 2026,
\newblock GitHub repository.

\end{thebibliography}

\end{document}